\documentclass[10pt]{article}
\usepackage[letterpaper]{geometry}
\usepackage{hicss}
\usepackage{times}
\usepackage[none]{hyphenat}
\usepackage{url}
\usepackage{latexsym}
\usepackage{amsmath}
\usepackage[frozencache,cachedir=.]{minted}
\usepackage{graphicx}
\graphicspath{{images/}}
\usepackage[
    style=ieee,
  ]{biblatex}
\usepackage{amssymb}
\usepackage{enumitem}

 \title{Quantifying Metrics for Wildfire Ignition Risk from \\[0.25em] Geographic Data in Power Shutoff Decision-Making}

 \author{Ryan Piansky \\
  Georgia Institute of Technology \\
  {\underline{rpiansky3@gatech.edu}} \\ \\
  Daniel K. Molzahn \\
  Georgia Institute of Technology \\
  {\underline{molzahn@gatech.edu} }
  \And
  Sofia Taylor \\
  University of Wisconsin--Madison \\
  {\underline{smtaylor8@wisc.edu} } \\ \\
  Line A. Roald\\
  University of Wisconsin--Madison \\
  {\underline{roald@wisc.edu} } 
  \And
  Noah Rhodes \\
  University of Wisconsin--Madison \\
  {\underline{nrhodes@wisc.edu} } 
  \\ \\
  Jean-Paul Watson\\
  Lawrence Livermore National Lab \\
  {\underline{watson61@llnl.gov} } \\ }

\date{}

\usepackage{float}
\floatstyle{ruled}
\newfloat{model}{thp}{lop}
\floatname{model}{Model}

\begin{document}
\maketitle
\begin{abstract}
\noindent Faults on power lines and other electric equipment are known to cause wildfire ignitions.
To mitigate the threat of wildfire ignitions from electric power infrastructure, many utilities preemptively de-energize power lines, which may result in power shutoffs. Data regarding wildfire ignition risks are key inputs for effective planning of power line de-energizations. However, there are multiple ways to formulate risk metrics that spatially aggregate wildfire risk map data, and there are different ways of leveraging this data to make decisions. The key contribution of this paper is to define and compare the results of employing six metrics for quantifying the wildfire ignition risks of power lines from risk maps, considering both threshold- and optimization-based methods for planning power line de-energizations. The numeric results use the California Test System (CATS), a large-scale synthetic grid model with power line corridors accurately representing California infrastructure, in combination with real Wildland Fire Potential Index data for a full year. This is the first application of optimal power shutoff planning on such a large and realistic test case. 
Our results show that the choice of risk metric significantly impacts the lines that are de-energized and the resulting load shed.
We find that the optimization-based method results in significantly less load shed than the threshold-based method while achieving the same risk reduction.
\end{abstract}

\subsubsection*{Keywords:}

Optimization, power shutoff, wildfire ignition risk.

\section{Introduction}
Power systems operators are increasingly concerned with the potential for electrical faults to ignite wildfires. In addition to an aging electric grid, both organic material buildup due to decades of fire suppression and climate change are intensifying fire risk conditions. In California, the land area burned due to wildfires is predicted to increase between 3\% and 52\% by 2050 based on a climate model ensemble \cite{Turco2023Anthropogenic}. Fires ignited from power equipment are common \cite{CPUCignitions} and tend to burn more area than wildfires ignited from other sources~\cite{syphard2015location}, likely because high winds and temperatures increase both fault probability and fire spread rate. 

To mitigate the risk of wildfire ignitions from power line faults, many utilities implement preemptive power shutoffs---called ``Public Safety Power Shutoffs'' (PSPS)---which involve selectively de-energizing certain power lines (via switching) to eliminate the possibility of these lines igniting fires. While this is an effective fire risk reduction strategy, it concurrently can lead to customer outages. Thus, operators face a complex environment in which they must avoid igniting wildfires by de-energizing lines while simultaneously ensuring reliable access to electricity. To better inform line de-energization planning, this paper characterizes how differences in wildfire ignition risk quantification can impact decisions in power grid operations. 

To prioritize risk mitigation actions, many utilities use threshold-based methods that de-energize lines whose associated risk values exceed a pre-determined threshold~\cite{huang2023review}.
Recently, researchers have proposed algorithms to better manage these trade-offs by considering both wildfire ignition risk and load shed due to power shutoffs when determining which lines to de-energize.
Reference~\cite{Rhodes2021Balancing} proposes an optimization model, referred to as the ``Optimal Power Shutoff'' (OPS) problem, to balance wildfire risk and load shed. Similar works and extensions are proposed for multi-period shutoff scheduling \cite{Astudillo2022Managing, LesageLandry2023Optimally}, security-constrained optimal power flow \cite{rhodes2023security}, stochastic unit commitment \cite{greenough2024wildfire}, alternative mitigation actions such as dynamic line ratings \cite{Tandon2021Motivating} and microgrid formation \cite{Taylor2023Managing}, power restoration \cite{Rhodes2023Cooptimization}, social equity considerations \cite{Kody2022Sharing}, and long-term investment planning \cite{Kody2022Optimizing, Bayani2023Resilient, piansky2024long}.
Machine learning techniques have also been applied to predict ignitions from power lines \cite{Yao2022Predicting} and to relate input wildfire scenarios and output mitigation strategies \cite{Tung2022Data}.
All of these models are sensitive to the specifics of wildfire risk parameters. Therefore, it is important to carefully consider how wildfire risk metrics for individual power lines are formulated.

As we will discuss in Section \ref{sec:metrics}, we define wildfire risk based on wildfire potential, or the likelihood for an ignition to spread into a large and devastating wildfire. Numerous approaches are available to quantify wildfire potential, leveraging a combination of surface meteorological measurements and satellite data.
This includes ``Fire Weather Watch'' and ``Red Flag Warning'' areas \cite{NWS2024Fire}, Significant Fire Potential Outlooks \cite{NIFC2024Outlooks}, and the Wildland Fire Potential Index (WFPI) \cite{USGS2024Wildland}.
Other tools simulate fire spread, such as FSim \cite{finney2011simulation} and FlamMap \cite{finney2006overview}. 
Another tool, Pyrecast, aggregates fire simulation results from millions of simulated ignition points to produce static maps of burned area risk \cite{pyregence}. 

\begin{figure}[thb]
    \centering
	\includegraphics[width=\linewidth]{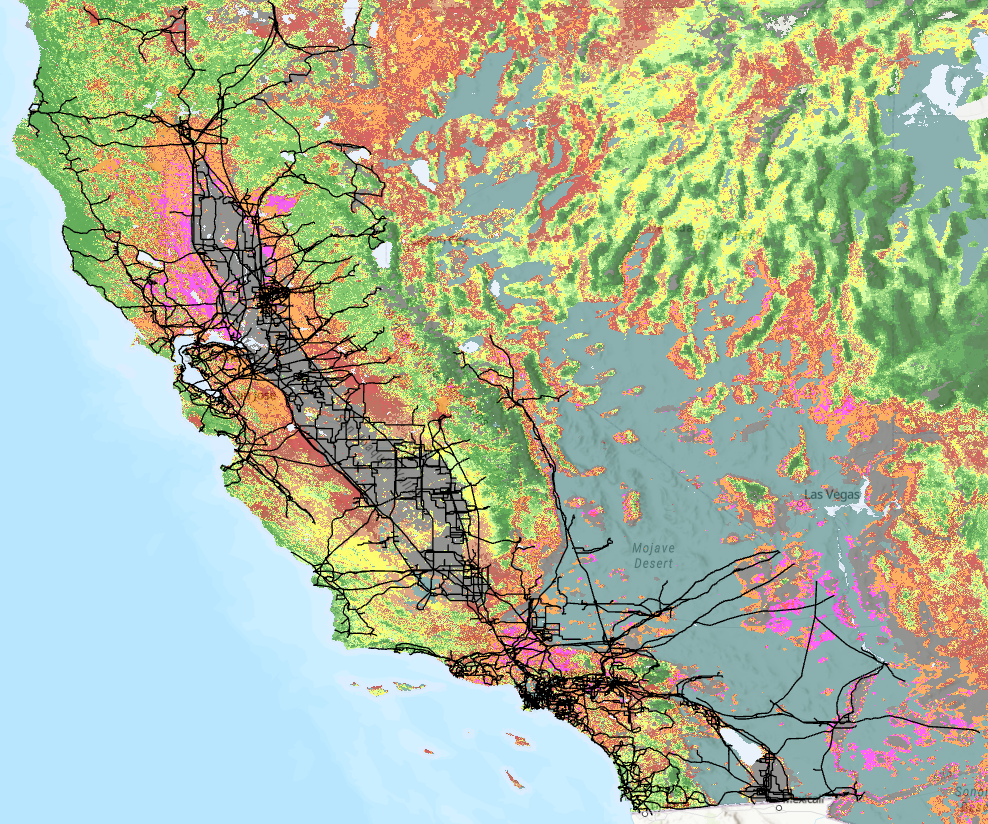}
	\caption{California's transmission line paths on a Wildland Fire Potential Index map for Oct.~26, 2020.}
	\label{fig:cats_wfpi_network}       
\end{figure}

We leverage the availability of high-fidelity, real-world wildfire risk maps to characterize the risk of energized lines.
WFPI and Pyrecast risk maps are good candidates for quantifying wildfire ignition risk from power equipment failure because of their temporal granularity (published daily and hourly, respectively), fine spatial granularity (1~square kilometer and 30 square meters, respectively), and the use of a range of potential levels (indices from 0 to 247 and the number of times each land area ``pixel'' is burned in the Monte Carlo sample set, respectively) rather than a binary ``threat'' categorization.
Researchers commonly overlay geospatial power grid data on top of such risk maps to obtain wildfire ignition risk values for individual power lines.
Fig.~\ref{fig:cats_wfpi_network} show California's transmission system superimposed on a WFPI map.


As seen in Fig.~\ref{fig:cats_wfpi_network}, power lines generally intersect multiple wildfire potential pixels or values, thus raising the question of how to appropriately aggregate risk values along the length of a line to obtain a single risk value for the line. 
Past works have used the maximum intersecting risk value \cite{Taylor2022Framework, Taylor2023Managing, Rhodes2023Cooptimization, greenough2024wildfire} or a sum of intersecting risk values (i.e., a cumulative metric) \cite{Kody2022Optimizing, Kody2022Sharing, pollack2024equitably}. 
However, it is not clear whether the maximum value accurately captures the risk of the entire line, as two lines with the same point-wise maximum value may have vastly different risk at other points; a decision-maker may be interested in prioritizing mitigation for the line with high risk along a significant fraction of the line length.  
Cumulative metrics, on the other hand, may not capture points of extreme risk, as the maximum value would.
This motivates alternate strategies of aggregating risk, such as computing the mean of all intersecting values or thresholding out low intersecting risk values. 
A comparison of maximum and cumulative wildfire risk metrics in~\cite{Taylor2022Framework} demonstrates that different risk metrics produce significantly different optimal capital investments for long-term wildfire risk mitigation. We note that the analysis in~\cite{Taylor2022Framework} has some limitations: the optimal investment model does not include a power flow model and the test network (RTS-GMLC \cite{RTSGMLC}) is relatively small and does not have realistic power line paths.

This paper addresses these limitations while also analyzing additional risk metrics. 
We focus on two challenges: (1) how to leverage wildfire risk maps to define the wildfire risk of power lines, 
and (2) how to leverage those metrics in power shutoff decision making for risk mitigation.
We compare the performance of six different risk metrics in two methods for determining line de-energizations to limit the wildfire risk in a power system: a threshold-based approach and an optimal power shutoff problem. We demonstrate results on the California Test System (CATS) \cite{taylor2023california}, a $\sim$9000-bus synthetic test system with transmission line paths that represent the actual geographical locations of transmission lines in California (as shown in Fig. \ref{fig:cats_wfpi_network}).

Our primary contributions are as follows:
\begin{itemize}[noitemsep, nolistsep]
\item  We compare the impacts of six different wildfire risk aggregation metrics on the resulting power shutoff plans. We find that these metrics result in significant differences among de-energization plans and load shed outcomes. To conduct this analysis, we processed two years of real-world wildfire potential maps to compute risk parameters for the transmission lines in the California Test System. This dataset is publicly available upon request.
\item We apply threshold- and optimization-based power shutoff planning to the California Test System, a synthetic yet highly realistic grid model with power line paths accurately representing California infrastructure.  We utilize demand and renewable generation availability profiles based on real hourly time-series data from the California Independent System Operator (CAISO) to simulate varying operating conditions\cite{taylor2023california}. This is the first application of the optimal power shutoff problem to such a large and realistic test case in the academic literature.
\item We find the that the optimization-based power shutoff method results in an approximately 80\% reduction of the load shed compared to the threshold-based method while maintaining the same overall wildfire ignition risk.
\end{itemize}

The remainder of the paper is organized as follows. Section~\ref{sec:metrics} introduces the six risk metrics we consider. Section~\ref{sec:shutoffs} details the methods for planning power grid shutoffs. Section~\ref{sec:results} describes our case study and presents computational results. Section~\ref{sec:conclusions} concludes the paper.

\section{Wildfire Risk Metrics} \label{sec:metrics}
In this section, we present six metrics that quantify the wildfire risk of individual power lines based on wildfire potential maps. To do this, we first discuss our definition of wildfire risk.

\subsection{Wildfire Risk Definition}
Consistent with the definitions in~\cite{Taylor2022Framework} and~\cite{Su2024Quasi}, we consider wildfire risk from power lines in terms of two components: fault probability and wildfire potential. 
An electrical fault occurs when an abnormal event (e.g., contact with vegetation or animals, conductor clashing, or downed power lines) results in current flow outside of a power line conductor. 
These events might involve arcing, sparks, and burning equipment or vegetation. 
Fault probability is influenced by factors specific to power systems in addition to weather conditions (e.g., wind) and vegetation factors. 
The likelihood of a fault is a function of the age and condition of infrastructure, right-of-way, line loading and sagging, and voltage level. 
The energy release associated with a fault can ignite a wildfire.
Wildfire potential, on the other hand, captures an ignition's impact, i.e., the subsequent potential for fire spread and intensity. This potential is dependent on factors that are not specific to power systems but rather is due to weather and vegetation conditions in the region surrounding an ignition. 

While many faults occur, the probability of a fault occurring at any particular time and place is small and difficult to assess. 
Reference~\cite{panossian2023power} proposes a risk metric that includes both fault probability and fire potential factors. 
Reference \cite{Bayani2023Quantifying} analyzes fault probability due to conductor clashing, comparing the use of a nonlinear model of conductor vibrational physics under wind forces with machine learning methods for fault prediction. 
Among other data requirements, these studies require information about distances between conductors and vegetation, the ground, and other conductors, which is not commonly available in models of real or synthetic power systems. Even utilities may not have this data, as line inspections are costly and time-consuming.
Thus, consistent with much of the literature on wildfire risk mitigation, in this paper we assume that the probability of a fault occurring is constant throughout the power system and define wildfire ignition risk based on wildfire potential only.  

\subsection{Aggregating Wildfire Potential Data}
We derive our risk metrics from  publicly available, real-world wildfire risk maps. 
We use WFPI maps from the U.S. Geological Survey \cite{USGS2024Wildland}. 
The WFPI geographically represents the relative potential for large fires and fire spread, and is published daily. 
The index is calculated at a spatial granularity of 1 km\textsuperscript{2} with a nominal range of 0-150.  The index is enhanced by wind speed and can exceed 150 with very high wind speeds.
Some land types such as desert and marshland do not have an associated value because they are considered ``unburnable''.
Agricultural land also does not have a risk value because the vegetation type and moisture levels change often and are not readily available. 

The wildfire risk metric for a power line is derived from the pixels of a wildfire risk map that a power line intersects. 
Lines typically cross many 1~km grid squares with potentially high variance in risk values, as shown in Fig.~\ref{fig:cats_wfpi_network}.  
Grid operators want to mitigate the risk of lines igniting fires, but are not able to de-energize just the high-risk segments of power lines. 
To assess the need for de-energization, we must aggregate the risk values that a power line passes through to obtain a  single risk value for the entire line.
Our primary objective is to analyze how the choice of risk aggregation metric impacts line de-energization decision-making.




We next define six power line wildfire risk metrics. The first three metrics are based on the mean, maximum, and cumulative values of the pixels the power lines intersect, respectively.  The last three metrics use an additional pre-processing step to only consider the pixels whose risk values are above a particular threshold.  
 Sections \ref{sec:baseline} and \ref{sec:hr_method} mathematically define the six risk metrics, while Section \ref{sec:discussion} discusses the intuition and real-world significance of the metrics.
\vspace{0.5em}

\subsection{Baseline Wildfire Risk Metrics}
\label{sec:baseline}
For a power line $\ell$, let $\mathcal{P}_{\ell}$ denote the set of pixel indices $p$ that the power line intersects.   
Let $\mathcal{R}_{\ell,d,p}$ denote the set of pixel risk values on day $d$ for line $\ell$.
For each line and on each day, we aggregate these values as the maximum, mean, or cumulative value. The formal definitions for each metric are provided below.

\paragraph{Maximum (MA) Metric}
assigns a risk value for each line equivalent to the maximum risk of any pixel that the line intersects:
\vspace{-0.5em}
\begin{equation}
R^{MA}_{\ell,d} = \max_{p \in \mathcal{P}_\ell}\, R_{\ell,d,p}.
\vspace{-1.0em}
\end{equation}

\paragraph{Mean (ME) Metric}
assigns a risk value that is the average of the pixels that the line intersects:
\vspace{-0.5em}
\begin{equation}
R^{ME}_{\ell,d} = \frac{\sum_{p \in \mathcal{P}_\ell} R_{\ell,d,p}}{|\mathcal{P}_\ell|}.
\vspace{-0.75em}
\end{equation}

\paragraph{Cumulative (CU) Metric}
assigns a risk that is the sum of the pixels that the line intersects:
\vspace{-1.0em}
\begin{equation}
R^{CU}_{\ell,d} = \sum_{p \in \mathcal{P}_\ell} R_{\ell,d,p}.
\vspace{-0.75em}
\end{equation}

\subsection{Pre-Processing Wildfire Risk}\label{sec:hr_method}
We now describe a method of pre-processing wildfire risk data to better identify the highest risk pixels before aggregating the risk values into a metric.
This method, adapted from \cite{pollack2024equitably}, uses a risk threshold, where pixel $p$ is removed from the set of pixels that a line traverses if it's risk $R_{\ell d p}$ is below the threshold.\footnote{We define this new set of pixels rather than set pixel values to zero for pixels below the threshold to better capture the intent of the high-risk mean metric.
With this definition, we will average only high-risk pixels rather than include zero values.}
The purpose of this processing is to ignore sections of lines where the risk is low, and only keep pixel risk values that are above this threshold.\footnote{We note this pre-processing threshold is separate from the threshold-based de--energization method discussed in Section~\ref{method_thresh}.}  

While there are many possible ways to compute the threshold value, we determine a value based on statistics of the wildfire risk values of a historical year. We take the set of all risk values $\mathcal{R}_{\ell,d,p}$ for all lines $\ell \in \mathcal{L}$, for all days $d \in D$, and for all pixel indices $p \in \mathcal{P}_\ell$.  We find the average risk value $\overline{r}$ and the standard deviation of risk values $\sigma$ of this set.  We define our threshold of interest as any risk value greater than one standard deviation above the mean.  The set of pixels above the threshold for each line on each day is defined as
\begin{equation}
    \mathcal{P}^{h}_{\ell,d} = \{p \in \mathcal{P}_\ell | R_{\ell,d,p} \geq \bar{r} + \sigma\},\;\;  \forall \ell \in \mathcal{L}, \forall d \in D.
\end{equation}
 We recompute the aggregation metrics with the high-risk thresholded risk values.

\paragraph{High-Risk Maximum (HRMA) Metric}
assigns a risk value equal to the maximum risk a line  intersects in the high-risk pixel set:
\vspace{-0.5em}
\begin{equation}
R^{HRMA}_{\ell,d} = \max_{p \in \mathcal{P}^{h}_{\ell,d}} R_{\ell,d,p}.
\vspace{-0.75em}
\end{equation}

\paragraph{High-Risk Mean (HRME) Metric}
assigns a risk value equal to the mean of the high-risk pixels that a line intersects:
\vspace{-0.6em}
\begin{equation}
R^{HRME}_{\ell,d} = \frac{\sum_{p \in \mathcal{P}^{h}_{\ell,d}} R_{\ell,d,p}}{|\mathcal{P}_{\ell}|}.
\vspace{-0.65em}
\end{equation}

\paragraph{High-Risk Cumulative (HRCU) Metric}
assigns a risk value equal to the sum of the high risk pixels that a line intersects:
\vspace{-0.5em}
\begin{equation}
R^{HRCU}_{\ell,d} = \sum_{p \in \mathcal{P}^{h}_{\ell,d}} R_{\ell,d,p}.
\vspace{-1.85em}
\end{equation}


\subsection{Risk Metric Discussion}
\label{sec:discussion}
The six risk metrics (MA, ME, CU, HRMA, HRME, and HRCU) each provide different incentives in a de-energization strategy, e.g. by focusing more or less on high versus average risk. We next provide a discussion and an illustrative example to demonstrate the similarities and differences between the metrics. 

\begin{figure}[thb]
    \centering
	\includegraphics[width=0.9\linewidth]{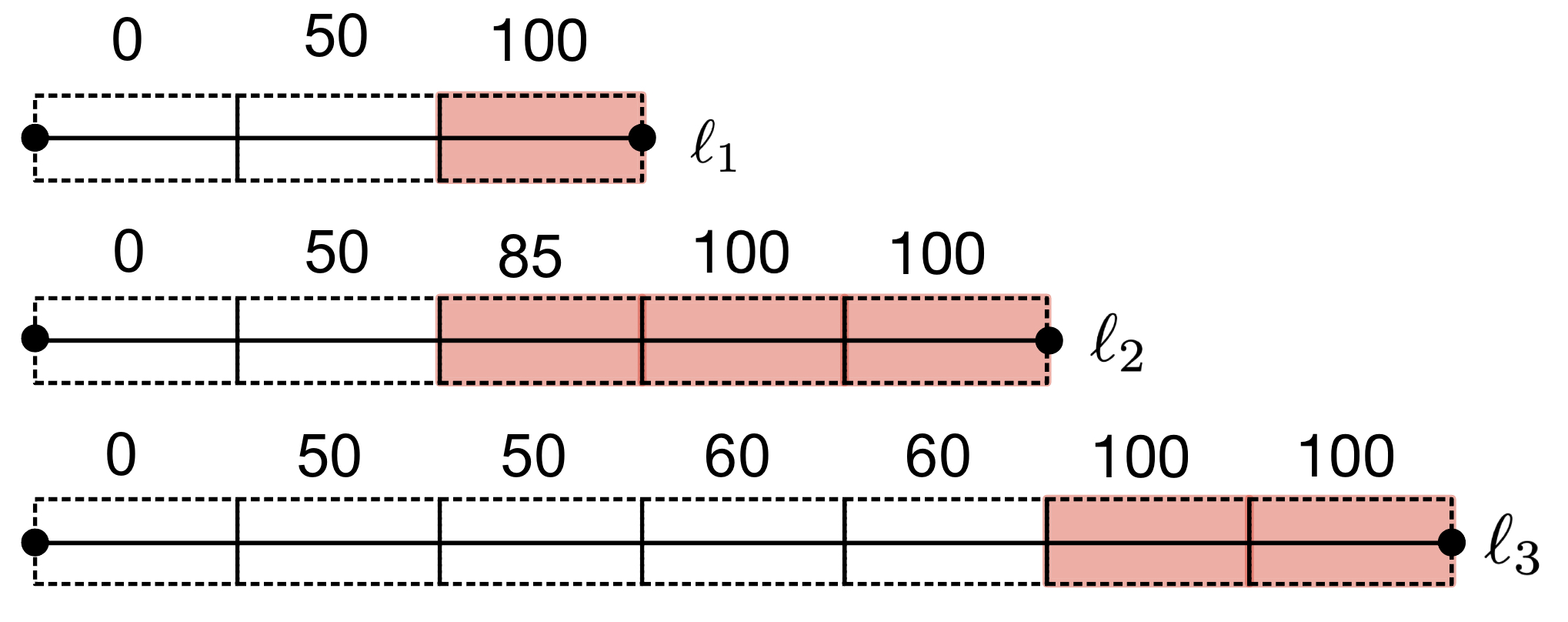}
	\caption{A conceptual example of three transmission lines. Boxes represent pixels that lines intersect, with their risk values shown above. Red pixels represent high-risk pixels in the set $\mathcal{P}^{h}_{\ell,d}$.}
	\label{fig:line_risk_examples}
\end{figure}

Fig. \ref{fig:line_risk_examples} shows three example lines of different length, traversing areas with different wildfire risk. Table 1 summarizes the wildfire risk metrics for each line.

\begin{table}[h]
\centering
\resizebox{\columnwidth}{!}{
\begin{tabular}{l|c|c|c|c|c|c}
          & MA   & HRMA & ME   & HRME & CU   & HRCU \\ \hline
$\ell_1$ & 100 & 100 & 50 & 33.3 & 150 & 100 \\ \hline
$\ell_2$ & 100 & 100 & 67 & 57 & 330 & 285 \\ \hline
$\ell_3$ & 100 & 100 & 60 & 28.6 & 420 & 200 \\ 
\end{tabular}
}
\caption{The risk values for the three lines in Fig.~\ref{fig:line_risk_examples} based on the six different risk metrics.}
\label{tab:example_risk_vals}
\end{table}

Selecting the maximum (MA) metric directly emphasizes the worst-case risk of a wildfire ignition along a corridor. 
This metric ignores the length of a power line, and therefore how many pixels it traverses and the variance of those risk values.  For instance, the shorter line $\ell_1$ in Fig.~\ref{fig:line_risk_examples} is treated the same as the longer line $\ell_3$, as both have a maximum intersecting risk value of 100.

The mean (ME) metric averages risk values across the length of a line. While the MA metric is concerned with only the point of worst-case risk, the ME metric considers the likelihood of any potential points of failure along a power line. This helps to identify lines that have a greater proportion of their length in high-risk areas. 
For example, line $\ell_2$ in Fig.~\ref{fig:line_risk_examples} has a relatively high mean risk value, as a relatively high proportion of this line intersects high-risk pixels.
Like MA, the ME metric does not capture the length factor of a power line's risk.

The cumulative (CU) metric sums all of the risk values along the length of the power line,  thus accounting for more possible instances of failures occurring along longer lines. This captures the intuition that a long line is more likely to start a fire than a short line (e.g., line $\ell_3$ is relatively long and has the highest CU risk of the lines in Fig.~\ref{fig:line_risk_examples}).
One limitation of this metric is that there may be cases where long lines pass through predominantly low-risk areas but nevertheless have high CU-based risk due to their long length.

The high-risk metrics (HRMA, HRME, and HRCU) threshold out low-risk pixel values. These reflect the intuition that wildfire ignitions at relatively low-risk locations are not likely to spread into devastating fires. The high-risk maximum risk (HRMA) metric is the same as the baseline maximum for any line that has a risk value above the risk threshold (e.g., we can see that all three lines in Fig.~\ref{fig:line_risk_examples} have equivalent MA and HRMA values).  
For any line whose maximum risk is below the risk threshold, the maximum value is zero rather than its baseline value. This removes any incentive to de-energize low- or moderate-risk power lines.

The high-risk mean (HRME) metric averages risk values across the length of a line after setting the low-risk pixels to a value of 0.
Compared to the baseline mean metric, this metric decreases the risk value of long lines with large sections in low-risk regions (e.g., observe that line $\ell_3$ in Fig.~\ref{fig:line_risk_examples} has a significantly lower HRME than ME).  

The high-risk cumulative (HRCU) metric sums all risk values above the threshold. Under this metric, a long line through a low-risk region will not be assigned a high risk value (note the lower HRCU than CU for line $\ell_3$), but lines with long stretches through high-risk regions will still have higher risk values compared to short lines through these regions.

In general, for each metric, a higher metric value for a line indicates that a fault-induced ignition on that line has a higher potential to spread into a large fire. Importantly, we can see that the method of risk aggregation (i.e., the choice of risk metric) significantly impacts which line(s) are considered to have the greatest wildfire potential. For example, in Fig.~\ref{fig:line_risk_examples}, line $\ell_2$ is riskiest based on the ME and HRCU metrics, but line $\ell_3$ is riskiest based on the CU metric.
For each metric, we expect that lines with high risk values would be better candidates for de-energization. In the remainder of this paper, we show how selecting different risk metrics impacts power shutoff planning decisions.

\section{Power Shutoff Decision-Making} \label{sec:shutoffs}
Given risk values for all lines in a power system, an operator can make a decision concerning which lines should be de-energized in order to decrease the overall potential for power infrastructure to ignite wildfires. We now assess two common analytic methods for determining how to implement power shutoffs: thresholding and optimal power shutoffs.

\subsection{Thresholding} \label{method_thresh}
\vspace{-0.8em}
Our first approach to planning PSPS events de-energizes any line that is considered ``risky'', i.e., any line that has a risk value that exceeds some predetermined level or threshold. This approach requires selection of a risk threshold that is low enough to turn off lines that are likely to ignite dangerous wildfires, yet high enough to avoid excessive de-energization and associated power outages.

We choose a threshold for each metric using statistical measures, specifically the 95\textsuperscript{th}-percentile of risk values (across all lines and all scenarios).
We de-energize lines that have risk values above this threshold on each day.
To determine how load is served under this shutoff plan, we solve a modified version of the optimization problem in Model \ref{model:ops}, which is described in detail below. Specifically, we fix the line status binary decision variables, $z^\ell$, to 0 for any de-energized line $\ell$.
The selected threshold results in lines being \emph{aggressively} de-energized, one to two orders of magnitude more than seen historically. This leads to large quantities of load shed, allowing for clear comparison of our metrics and de-energization methods.

\subsection{Optimal Power Shutoff}\label{sec:ops}
The optimal power shutoff (OPS) problem is an optimization problem that determines steady-state operations decisions (including generator outputs, line flows, loads served, and voltage angles) as well as binary line de-energization decisions in a way that balances wildfire risk mitigation with load that is shed due to those de-energizations. The OPS was first proposed in \cite{Rhodes2021Balancing} and is studied and extended in \cite{Astudillo2022Managing, LesageLandry2023Optimally, rhodes2023security,greenough2024wildfire,Tandon2021Motivating,Taylor2023Managing,Rhodes2023Cooptimization,Kody2022Sharing,Kody2022Optimizing,Bayani2023Resilient,piansky2024long,pollack2024equitably}.

There are several ways to formulate the risk and load shed mitigation strategies in the OPS problem. For example, we can formulate a multi-objective problem that minimizes wildfire risk and load shed, as in \cite{Rhodes2021Balancing, Kody2022Optimizing, Kody2022Sharing}, or we can minimize wildfire risk while constraining load shed to some acceptable level as in \cite{rhodes2023security} (or vice versa as in \cite{pollack2024equitably}). In this paper, given we are interested in comparing risk metrics, we choose to minimize load shed while constraining wildfire risk. This allows us to maintain a safe level of wildfire risk in the network while optimizing to reduce negative impact on loads.
The OPS formulation is outlined in Model~\ref{model:ops}.

\begin{model}[t]
\caption{Optimal Power Shutoff (OPS)}
\label{model:ops}
\begin{subequations}
\vspace{-0.2cm}
\begin{align}
& \mbox{\textbf{min}} \  \sum_{t \in  \mathcal{T}}\sum_{n \in \mathcal{N}} p_{ls,t}^n + \epsilon_{\text{switch}}\sum_{\ell \in \mathcal{L}^{\text{switch}}} (1-z^\ell) \label{eq:obj} 
\\
& \mbox{\textbf{s.t.}}  \quad \forall t \in \mathcal{T}, \nonumber \\
& \underline{p}_{g}^i \leqslant p_{g,t}^i \leqslant \overline{p}_{g}^i 
&& \hspace{-5.1em} \forall i \in \mathcal{G} \label{subeq: gen limits} 
\\
& 0 \leqslant p_{ls,t}^n \leqslant p_{d,t}^n 
&& \hspace{-5.1em} \forall n \in \mathcal{N} \label{subeq: loadshed limits} 
\\ 
& -\overline{f}^\ell z^\ell \leqslant f^\ell_{t} \leqslant \overline{f}^\ell z^\ell 
&& \hspace{-5.1em} \forall \ell \in \mathcal{L}^{\text{switch}} \label{subeq: power flow limits switching}
\\
& -\overline{f}^\ell \leqslant f^\ell_{t} \leqslant \overline{f}^\ell 
&& \hspace{-5.1em} \forall \ell \in \mathcal{L} \setminus \mathcal{L}^{\text{switch}} \label{subeq: power flow limits} 
\\
& \theta^{n^{\ell, \text{fr}}}_{t} \!\!\!\! - \theta^{n^{\ell, \text{to}}}_{t}  \!\!\!\! \geqslant \underline{\delta}^\ell z^\ell \!\! + \underline{M}(1\!-\!z^\ell) 
&& \hspace{-5.1em} \forall \ell \in \mathcal{L}^{\text{switch}} \label{subeq: voltage angle switching 1} 
\\
& \theta^{n^{\ell, \text{fr}}}_{t} \!\!\!\! - \theta^{n^{\ell, \text{to}}}_{t} \!\!\!\!  \leqslant  \overline{\delta}^\ell z^\ell \!\! + \overline{M}(1\!-\!z^\ell)
&& \hspace{-5.1em} \forall \ell \in \mathcal{L}^{\text{switch}} \label{subeq: voltage angle switching 2} 
\\
& \underline{\delta}^\ell  \leqslant \theta^{n^{\ell, \text{fr}}}_{t} \!\!\!\! - \theta^{n^{\ell, \text{to}}}_{t} \leqslant \overline{\delta}^\ell 
&& \hspace{-5.1em} \forall \ell \in \mathcal{L}\setminus\mathcal{L}^{\text{switch}} \label{subeq: voltage angle} 
\\
& f_{t}^\ell \! \geqslant \!\! -b^\ell(\theta^{n^{\ell, \text{fr}}}_{t} \!\!\!\!\!\! - \theta^{n^{\ell, \text{to}}}_{t}) \! + \! |b^\ell|\underline{M}(1 \! - \! z^\ell) 
&& \hspace{-3.15em} 
\forall \ell \in \mathcal{L}^{\text{switch}} \label{subeq: power flow switching 1} 
\\
& f_{t}^\ell \! \leqslant \!\! -b^\ell(\theta^{n^{\ell, \text{fr}}}_{t} \!\!\!\!\!\! - \theta^{n^{\ell, \text{to}}}_{t}) \! + \! |b^\ell|\overline{M}(1 \!- \!z^\ell) 
&& \hspace{-3.15em} 
\forall \ell \in \mathcal{L}^{\text{switch}} \label{subeq: power flow switching 2} 
\\
& f_t^\ell \geqslant -b^\ell(\theta^{n^{\ell, \text{fr}}}_{t} \!\!\!\! - \theta^{n^{\ell, \text{to}}}_{t}) 
&& \hspace{-5.1em}  \forall \ell \in \mathcal{L}\setminus\mathcal{L}^{\text{switch}} \label{subeq: power flow 1}
\\
& f_t^\ell \leqslant  -b^\ell(\theta^{n^{\ell, \text{fr}}}_{t} \!\!\!\! - \theta^{n^{\ell, \text{to}}}_{t}) 
&& \hspace{-5.1em} \forall \ell \in \mathcal{L}\setminus\mathcal{L}^{\text{switch}} \label{subeq: power flow 2}
\\
& \! \sum_{\ell \in \mathcal{L}^{n, \text{fr}}} \!\!\!\! f^\ell_{t} \! - \!\!\!\! \sum_{\ell \in \mathcal{L}^{n, \text{to}}} \!\!\!\! f^\ell_{t} \! = \!\!\!  \sum_{i \in \mathcal{G}^n} \!\! p_{g,t}^i \! - \! p_{d,t}^n \! + \! p_{ls,t}^n 
&& \hspace{-2.5em} 
\forall n \in \mathcal{N} \label{subeq: power balance}
\\
& \! \sum_{\ell \in \mathcal{L}^{\text{switch}}} z^{\ell} R_{\ell,d} \leq R^{PSPS}_d && \hspace{-5.1em} \forall \ell \in \mathcal{L}^{\text{switch}}. \label{subeq: risk threshold}
\end{align}
\end{subequations}
\end{model}

Equation \eqref{subeq: gen limits} enforces lower ($\underline{p}_{g}^i$) and upper ($\overline{p}_{g}^i$) generation limits for power generation ($p_{g,t}^i$) at all generators $i \in \mathcal{G}$ at all times $t \in \mathcal{T}$. Equation \eqref{subeq: loadshed limits} constrains any load shedding ($p_{ls,t}^n$) to be nonnegative and less than the power demand ($p_{d,t}^n$) at that time at each bus $n \in \mathcal{N}$. Equations~\eqref{subeq: power flow limits switching} and \eqref{subeq: power flow limits} enforce line flow ($f^\ell_{t}$) to be within lower and upper limits ($\overline{f}^\ell$) in accordance with the energization status, $z^\ell$, for lines $\ell \in \mathcal{L}$. For lines not in the switchable set $\mathcal{L}^{\text{switch}}$, note that $z^\ell$ is required to be set to one (indicating that the line is energized) instead of being a binary decision variable, thus reducing \eqref{subeq: power flow limits switching} to \eqref{subeq: power flow limits}. Lines with zero risk are excluded from the switchable set $\mathcal{L}^{\text{switch}}$. 

Equations \eqref{subeq: voltage angle switching 1}, \eqref{subeq: voltage angle switching 2}, and \eqref{subeq: voltage angle} constrain the difference in voltage angle across a line $\ell$ from the from-bus $n^{\ell,fr}$ to the to-bus $n^{\ell,to}$ at each time step to be within the lower ($\underline{\delta}^\ell$) and upper ($\overline{\delta}^\ell$) limits in accordance with the line energization status. Again, if a line is not considered switchable, the energization status $z^\ell$ is set to one, thus reducing \eqref{subeq: voltage angle switching 1} and \eqref{subeq: voltage angle switching 2} to \eqref{subeq: voltage angle}. 

Equations \eqref{subeq: power flow switching 1}, \eqref{subeq: power flow switching 2}, \eqref{subeq: power flow 1}, and \eqref{subeq: power flow 2} model the DC power flow approximation with line energization status. Equations \eqref{subeq: voltage angle switching 1}, \eqref{subeq: voltage angle switching 2}, \eqref{subeq: power flow switching 1}, and \eqref{subeq: power flow switching 2} utilize big-M constants to allow for voltage angle differences to be unconstrained across de-energized lines, with $\overline{M}$ and $\underline{M}$ set to $2\pi$ and $-2\pi$ respectively for results shown in this paper. Note that more sophisticated methods for selecting big-M values could be used, such as those in \cite{pineda2023tight}.
Equation \eqref{subeq: power balance} ensures power balance at all buses in the network. Equation \eqref{subeq: risk threshold} restricts the total risk on the network to be below a value, $R^{PSPS}$, defined by the total remaining risk on the network resulting from the thresholded lines on the same day (see Section \ref{method_thresh}). Here, $R_{\ell,d}$ represents the risk on line $\ell$ based on the given metric. The objective \eqref{eq:obj} minimizes total load shed in the network with an associated penalty term on the number of de-energized lines. For the numerical results in Section \ref{sec:results}, $\epsilon_{\text{switch}}$ is set to 0.01 to avoid the line switching decisions dominating the objective.

\section{Test Case Results} \label{sec:results}

We compare the performance of the six wildfire risk metrics and two shutoff methods utilizing CATS  \cite{taylor2023california}, a $\sim$9000-bus, $\sim$11000-line test system with transmission line corridors that reflect the actual grid in California, but with synthetic parameters so as not to reveal any critical information about the real grid. The geographic realism and large scale of CATS make it a compelling test case. However, we note that the results derived here are not necessarily an indication of how California's actual power grid operates. We use 2019 WFPI daily risk data to determine high-risk pixel thresholds, and study de-energization plans using 2020 WFPI data. All models were implemented using Julia 1.9.2~\cite{bezanson2017julia} and solved using Gurobi~10.0.1~\cite{gurobi}.



To reduce solve times (due to the scale of CATS), we take a number of approaches. First, when finding optimal switching decisions, we warm-start the binary variables in the problem with the status of lines de-energized under the corresponding thresholded case. We also reduce the number of binary variables by only allowing lines with non-zero wildfire risk to be included in the switchable line set, $\mathcal{L}^{\text{switch}}$. Additionally, we relax the lower bound of generators to be 0 p.u.~to avoid binary decision variables associated with generator on/off statuses.
Finally, to reduce the problem size, we make optimal switching decisions for the entire day based on the worst-case hour of the day. We define the worst-case hour as the hour with the most load shed from thresholded de-energization decisions from the same day with the same metric. After optimal de-energization decisions are made on the given hour, these decisions are fixed for the full 24-hour period, with hourly load shed and operational decisions found for the full day. 


\subsection{De-energization Decisions based on Thresholding}\label{sec:results thresh}

\begin{figure}[h]
    \centering
    \includegraphics[width=.99\linewidth]{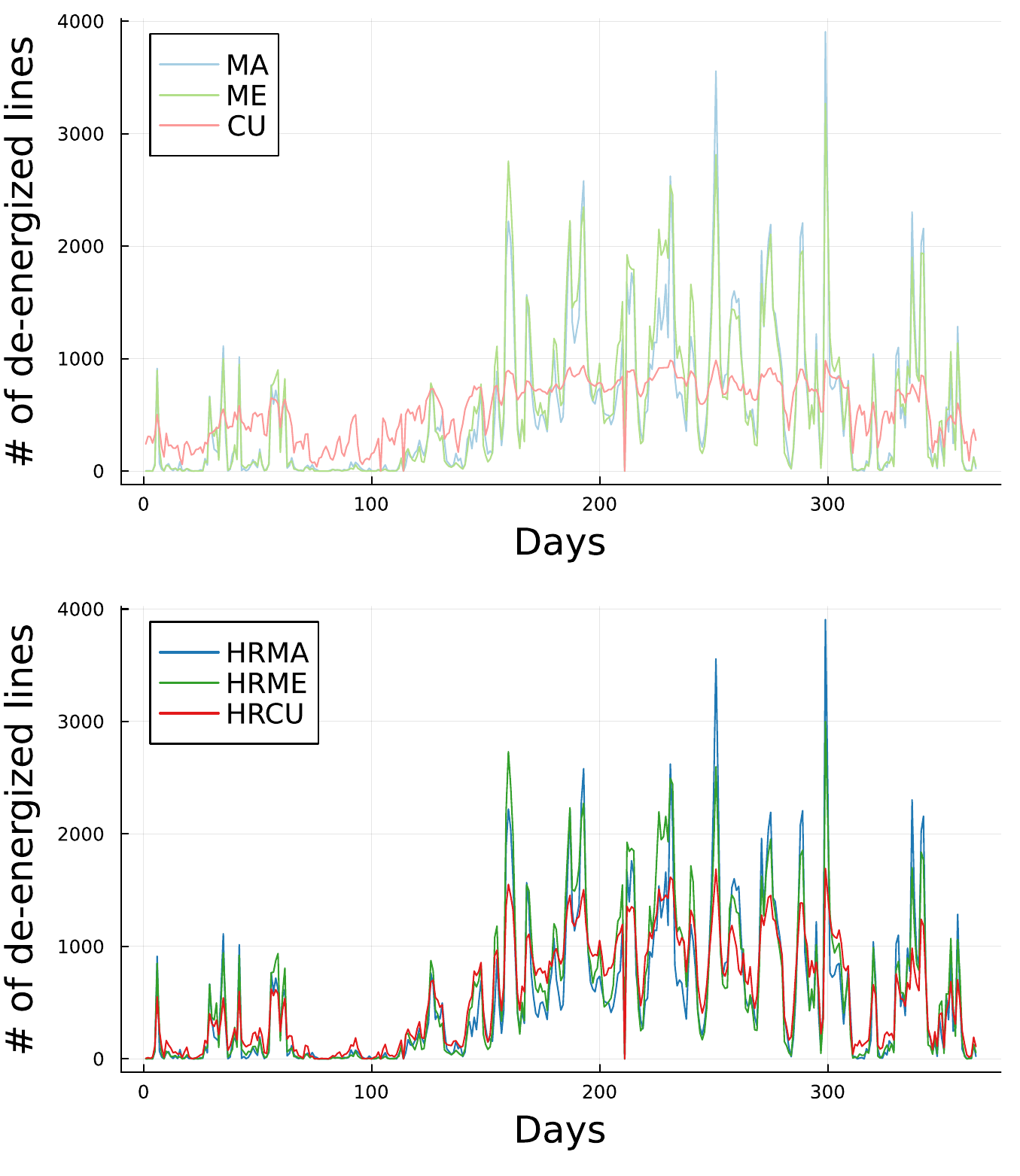}
   \caption{The number of lines de-energized daily by metric under the 95\textsuperscript{th}-percentile thresholding method.}
    \label{fig:thresholded_line_decisions}
\end{figure}

Thresholding decisions differ between the normal and high-risk methodologies. As shown in Fig.~\ref{fig:thresholded_line_decisions}, each method provides different line de-energization outcomes throughout the year. Both the CU and HRCU see a more uniform number of de-energized lines throughout the year while MA, ME, HRMA, and HRME all see more variation in numbers of de-energized lines. Fig.~\ref{fig:thresholded_line_decisions} also shows that the the number of de-energized lines resulting from metrics with the high-risk pixel threshold (HRMA, HRME, HRCU) are  more closely aligned throughout the year than those from the metrics which use the raw wildfire data (MA, ME, CU).

\begin{figure}[h]
\centering
\includegraphics[width=.99\linewidth]{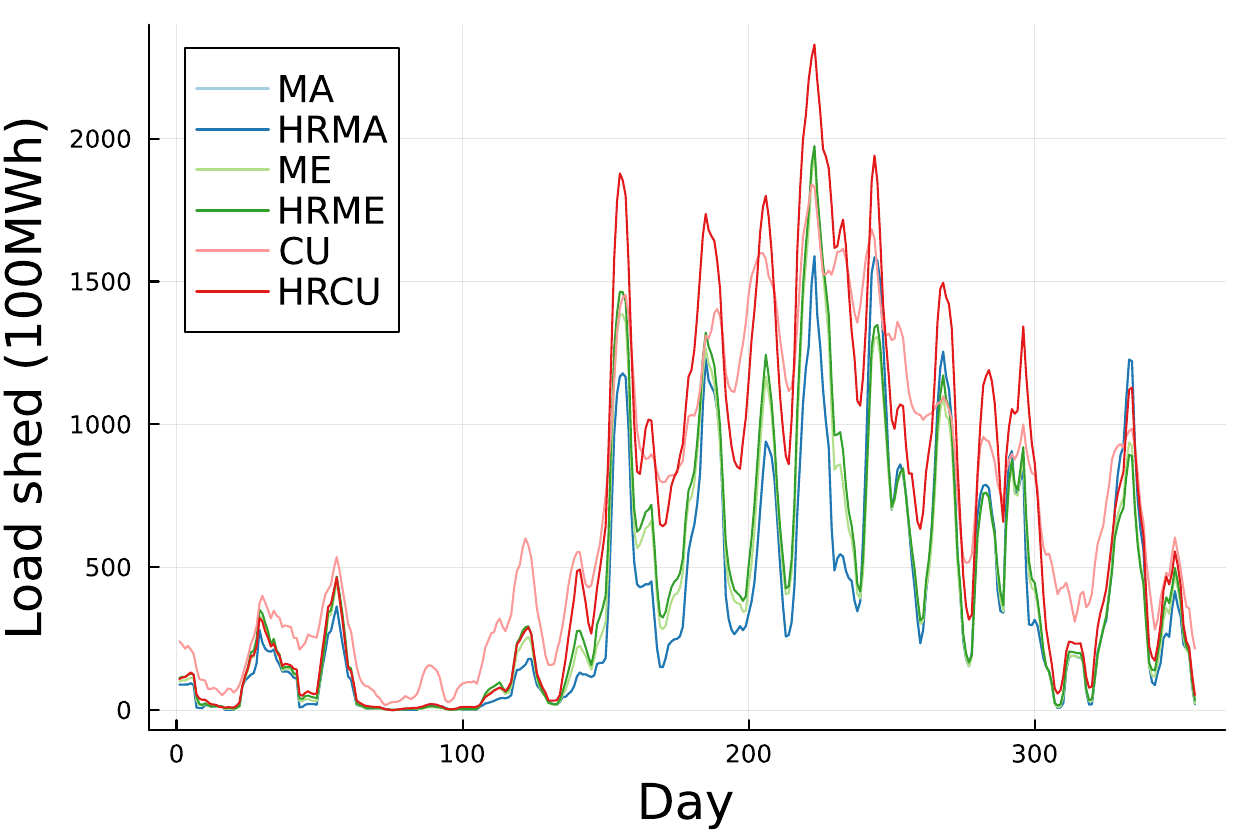}  
\caption{Seven-day rolling average load shed from thresholded line de-energizations.}
\label{fig:thresh_avg_ls}
\end{figure}

The first row of Table~\ref{tab:unique_off_decisions} shows the number of unique lines de-energized by the thresholded method in 2020. We note that while the MA and ME metrics de-energize nearly the same number of unique lines as the HRMA and HRME metrics, the HRCU metric de-energizes nearly twice as many unique lines as the CU metric over the span of the year. The resulting daily load shed for each metric is shown in Fig.~\ref{fig:thresh_avg_ls}. While the HRCU metric de-energizes fewer lines than the max- or mean- derived metrics, it results in the most load shed throughout the year. We also note that the MA and HRMA metrics de-energize the same number of unique lines (and the most by the thresholding method) but result in less load shed compared to other metrics for most of the year.

\subsection{De-energization Decisions based on OPS}

\begin{figure}[h]
    \centering
    \includegraphics[width=.99\linewidth]{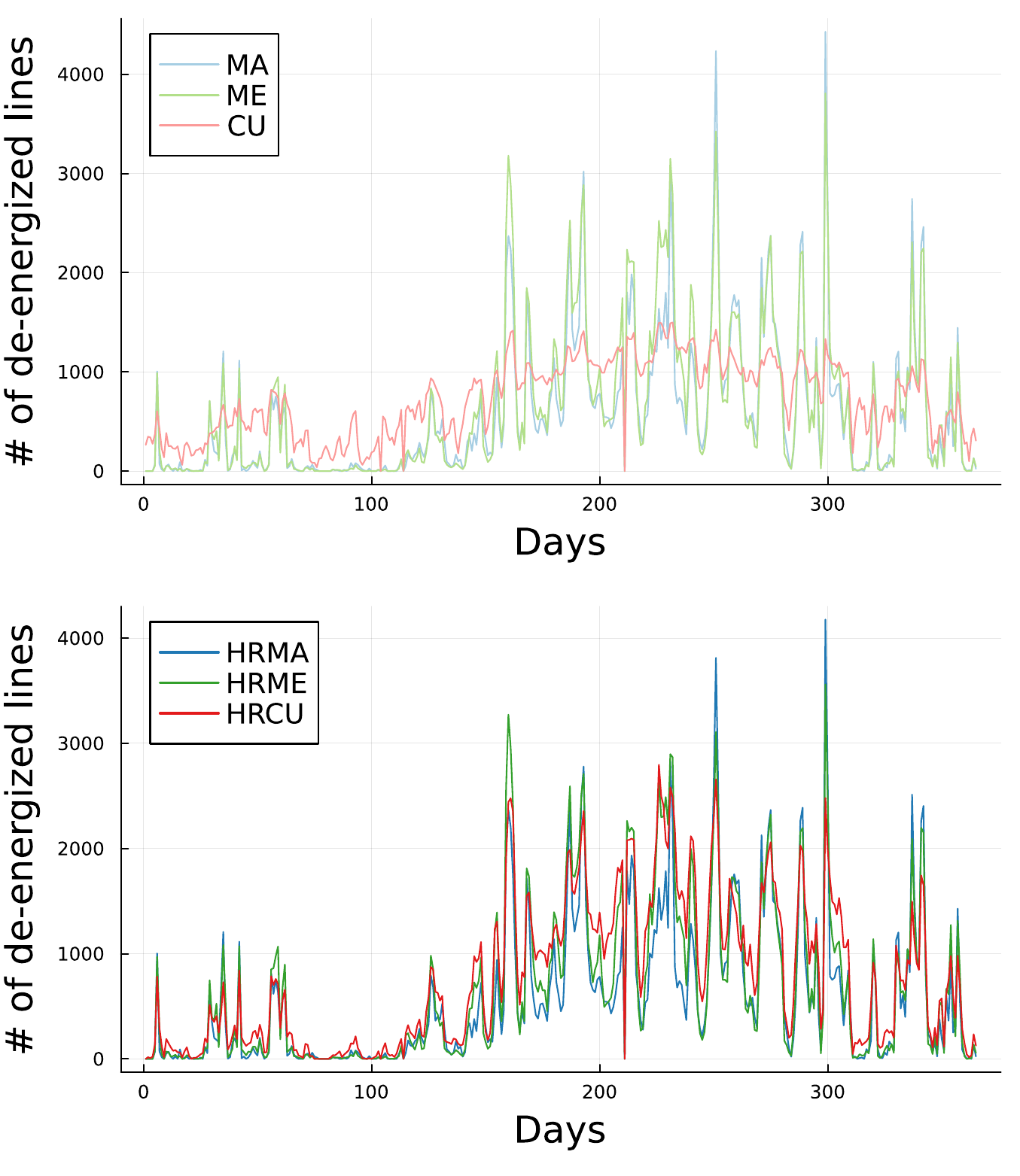}
    \caption{The number of lines de-energized daily, by metric, under the optimal power shutoff method.}
    \label{fig:optimal_line_decisions}
\end{figure}

\begin{figure}[t]
\centering
\includegraphics[width=.99\linewidth]{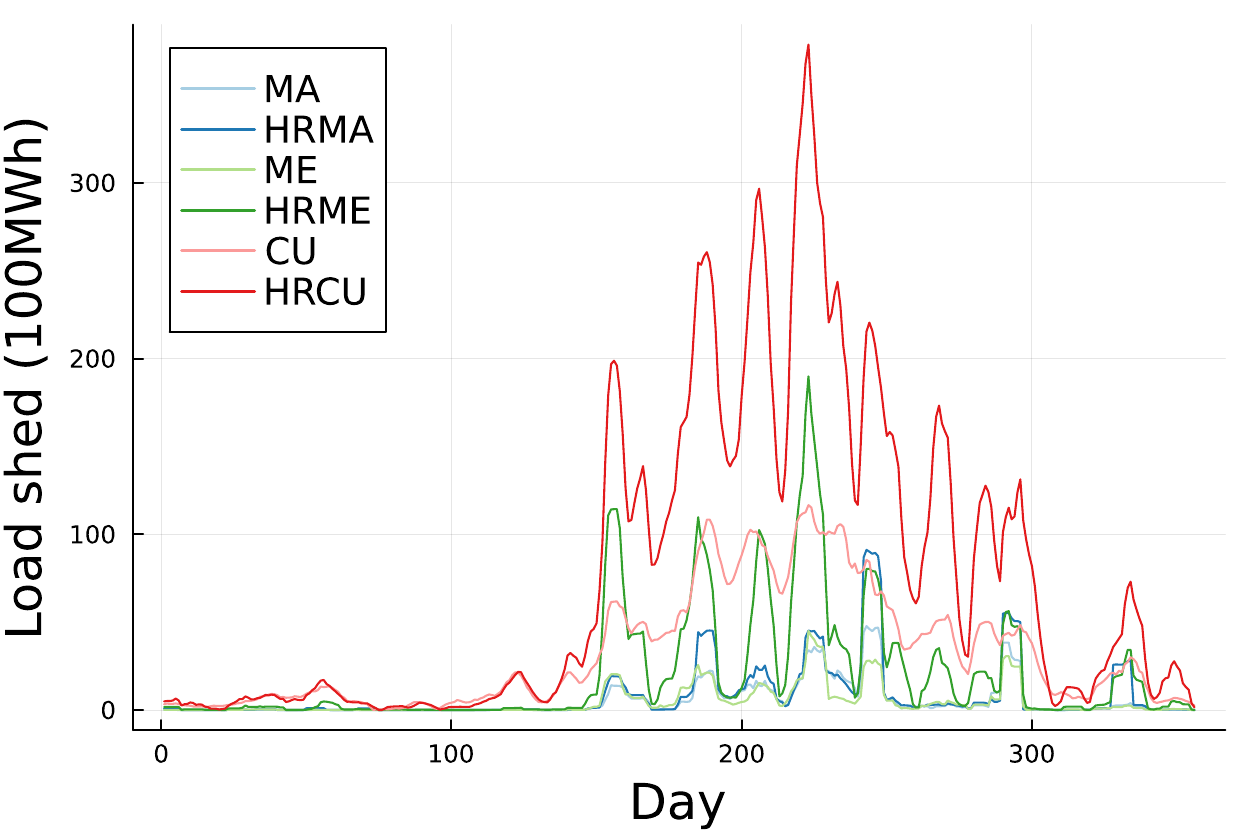}  
\caption{Seven-day rolling average load shed from optimal power shutoff line de-energization decisions.}
\label{fig:opt_avg_ls}
\end{figure}

When looking at the number of optimally de-energized lines in Fig.~\ref{fig:optimal_line_decisions}, we see that the overall quantity of lines being de-energized is similar to the thresholded method. However, the optimal case de-energizes slightly more lines across all metrics. This is likely due to more lower-risk lines being de-energized to allow energization of some high-risk lines that are crucial for power delivery. By strategically de-energizing more low-risk lines, the overall risk in the network can be maintained while reducing the amount of load shed. Fig.~\ref{fig:opt_avg_ls} shows that the optimal line switching decisions achieve approximately 20\% of the load shed resulting from the thresholded method. We note that the scale of the y-axis differs from that in Fig.~\ref{fig:thresh_avg_ls} to allow for variations in the load shed to be visible. Similar to the thresholded method, we again see that the CU and HRCU methods, while de-energizing fewer lines, result in larger amounts of load shed. 

\subsection{Comparing Thresholding and Optimal Decisions}

\begin{figure}[t]
\centering
\includegraphics[width=.99\linewidth]{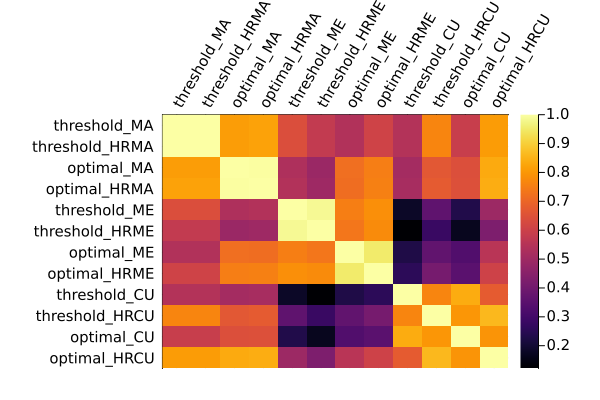}  
\caption{This heatmap shows how similar the line de-energization decisions are between the thresholded method and optimal methods with each metric over the full year. A value of 1.0 (yellow) indicates each line was de-energized the same number of times over the year while a value of 0.0 (black) indicates none of the same lines were de-energized over the year.}
\label{fig:corr_heatmap}
\end{figure}

While both thresholded and optimal de-energization decisions result in a similar number of lines being switched off, they do not select the same lines to be turned off throughout the year. Fig.~\ref{fig:corr_heatmap} shows the similarity between decisions made by each method/metric pairing. For each method/metric pair, we sum the number of times each individual line is de-energized. We then normalize each of these vectors and take the dot product. Thus, a value of 1.0 in the heat map indicates that each line $\ell$ was de-energized the same number of times throughout the year. 

Fig.~\ref{fig:corr_heatmap} shows that thresholded MA and HRMA metrics produce the same results, and that the optimal MA and HRMA metrics produce nearly identical results. We also see that the decisions made by the ME or HRME metrics have very weak correlations with those made by the CU or HRCU metrics. Also note that the CU and HRCU metrics result in dissimilardecisions and optimal load shed. This might be caused by the HRCU method de-energizing nearly 60\% more lines compared to the CU method in the optimal case (as shown in Table~\ref{tab:unique_off_decisions}). 

\begin{table}[h]
\centering
\resizebox{\columnwidth}{!}{
\begin{tabular}{l|c|c|c|c|c|c}
          & MA   & HRMA & ME   & HRME & CU   & HRCU \\ \hline
Threshold & 6376 & 6376 & 5204 & 5005 & 1206 & 2505 \\ \hline
OPS       & 6276 & 6398 & 5564 & 5859 & 3568 & 5649 \\ 
\end{tabular}
}
\caption{The number of unique lines de-energized in 2020 for each metric in the threshold- and optimization-based methods.}
\label{tab:unique_off_decisions}
\end{table}

\section{Conclusions} \label{sec:conclusions}

This paper characterizes the impacts of different approaches to aggregating wildfire risk data for power lines using a realistic large-scale synthetic power system. We define and compare six wildfire risk aggregation metrics using two methods for making line de-energization decisions, thresholding and the optimal power shutoff. 
We find that, compared to thresholding, the optimal power shutoff formulation drastically reduces load shed, despite a similar extent of line de-energizations and similar overall network risk.
The numerical results also clearly show that the choice of metric significantly alters the de-energization decisions and associated load shed. If we had found that all six metrics produced similar de-energization results, then the choice of risk quantification metric in future analyses would not matter.~However, since the results significantly differ, modelers and decision-makers should be careful when choosing how to aggregate risk. 

Based on our results alone, it is not clear which risk metric or method is the most effective or fair to use in planning pre-emptive deenergizations. To determine this, more research is needed. For example, high-fidelity fire simulations could achieve a thorough analysis of the impact of ignitions that would occur (or be avoided), which could help determine if focusing on high risk locations or average risk across the line span is more important. Further, while optimization-based methods achieve lower load shed at comparable wildfire risk compared with the threshold-based method, they may leave certain high risk areas unaddressed to avoid load shedding in further away locations. Understanding the effects of trade-offs and metrics on both local communities and the overall grid is a topic that deserves more attention in future work.

Future work will also compare modeled de-energization decisions with those made historically.
In addition, extensions to consider alternative modeling formulations would be valuable, such as a security-constrained optimal power shutoff, AC optimal power shutoff, and multi-period optimization. 

\vspace{-0.6em}
\section*{Acknowledgements}
\vspace{-0.8em}
\noindent This work was performed in part under the auspices of the U.S. Dept.~of Energy (DOE) by Lawrence Livermore National Lab under Contract DE-AC52-07NA27344 and was supported by the LLNL LDRD Program under Project 22-SI-008 and the DOE Office of Electricity's Advanced Grid Modeling program.~D.K.~Molzahn and R.~Pianksy acknowledge support from the NSF AI Institute for Advances in Optimization (AI4OPT), \#2112533.~N.~Rhodes and L.A.~Roald acknowledge support from the National Science Foundation (NSF) under the NSF CAREER award \#2045860.~S.~Taylor is supported by the NSF Graduate Research Fellowship under Grant \#DGE-1747503.~We thank Gurobi for providing the use of academic licenses.


\printbibliography

\end{document}